\documentclass[graybox]{svmult}

\usepackage{mathptmx}       
\usepackage{helvet}         
\usepackage{courier}        
\usepackage[misc]{ifsym}
\usepackage{makeidx}         
\usepackage{graphicx}        
\usepackage{multicol}        
\usepackage[bottom]{footmisc}
\usepackage{subfigure}
\usepackage{multirow}
\usepackage[table,xcdraw]{xcolor}
\usepackage{url}

\makeindex             

\newcommand{\expr}[1]{Exp.~#1}
\newcommand{\proc}[1]{Proc.~#1}

\begin{document}

\title*{Micro and Macro Pedestrian Dynamics in Counterflow: the Impact of Social Groups}
\titlerunning{Micro and Macro Pedestrian Dynamics in Counterflow} 
\author{Luca Crociani, Andrea Gorrini, Claudio Feliciani, Giuseppe Vizzari, Katsuhiro Nishinari, Stefania Bandini}
\authorrunning{Crociani, L., Gorrini, A., Feliciani, C., Vizzari, G., Nishinari, K., Bandini, S.}
\institute{Luca Crociani(\Letter) \and Andrea Gorrini \and Giuseppe Vizzari \and Stefania Bandini \at CSAI research center, University of Milano-Bicocca, Milan, Italy.\\\email{{name.surname}@disco.unimib.it}
\and Claudio Feliciani \and Katsuhiro Nishinari \and Stefania Bandini \at The University of Tokyo, Tokyo, Japan.
\\\email{feliciani@jamology.rcast.u-tokyo.ac.jp}
\and Katsuhiro Nishinari
\\\email{tknishi@mail.ecc.u-tokyo.ac.jp}
}
%
%
\maketitle

\abstract{Although it is widely recognised that the presence of groups influences microscopic and aggregated pedestrian dynamics, a precise characterisation of the phenomenon still calls for evidences and insights. 
The present paper describes micro and macro level original analyses on data characterising pedestrian behaviour in presence of counter-flows and grouping, in particular dyads, acquired through controlled experiments. Results suggest that the presence of dyads and their tendency to walk in a line-abreast formation influences the formation of lanes and, in turn, aggregated observables, such as overall specific flow.}

\section{Introduction}
\label{sec:1}
Research on pedestrian dynamics has systematically analysed the influence of group behaviour only in the last years (see, e.g.,~\cite{TheraulazGroup,DBLP:journals/prl/BandiniGV14,VonKruchten2017}): although observations and experiments agree on some aggregated and microscopic effects of the presence of groups (e.g. group members walk slower than individuals), there is still need for additional insights, for instance on the spatial patterns assumed by groups in their movement and in general on the interaction among different factors influencing overall pedestrian dynamics (e.g.~do obstacles still make egress from a room smoother in presence of groups?). Models incorporating mechanisms reproducing the cohesion of group members, in fact, are just partially able to reproduce overall phenomena related to the presence of groups in the simulated population of pedestrians (see, e.g.,~\cite{ITSC2014-groups} in which groups preserve their cohesion and they move slower than individuals) and they would benefit from additional insights on how members manage their movements balancing (for instance) goal orientation, tendency to stay close to other members, opportunities offered by the presence of lanes.

In this framework, the present work discusses results of experiments carried out to investigate the potentially combined impact of counter-flow situations~\cite{Zhang2012} and grouping \cite{zanlungo2014potential}. Experiment 1~\cite{feliciani2016empirical} tested the impact of four different configurations of counter-flow in a corridor setting (from uni-directional to fully balanced bi-directional flow). Experiment 2~\cite{gorrini2016social} replicated the same procedures and about half of participants were paired to compose dyads (the simplest and most frequent type of group), asking them to walk close to their companion. In the following, both the experimental procedures and the achieved results will be presented in details.

\section{Description of Experiments}
\label{sec:exp}

\begin{figure}[t]
\begin{center}
\includegraphics[width=.99\textwidth]{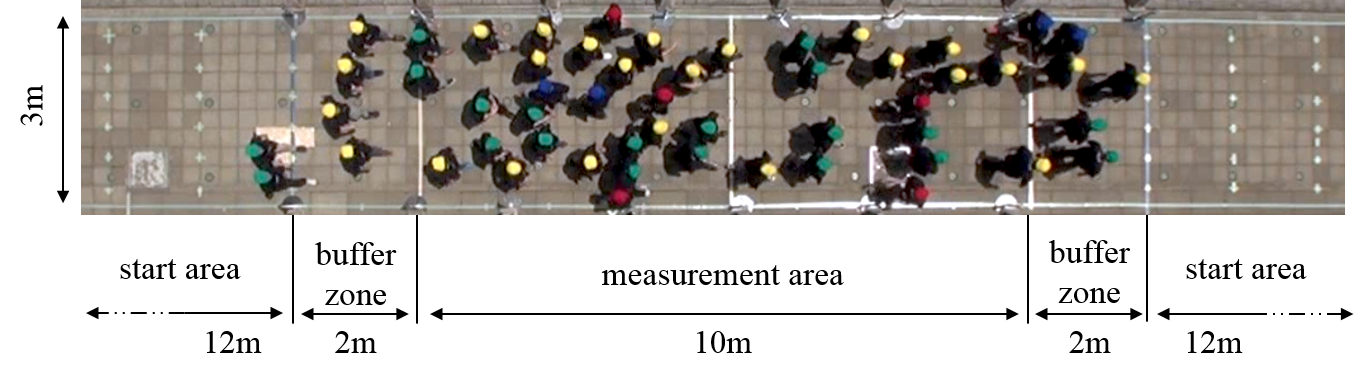}
\caption{A screenshot from the video of the experiments and a schematic representation of the setting.}\label{fig:experiments-setting}
\end{center}
\end{figure}

The two experiments have been performed on June 13, 2015 at the Research Center for Advanced Science and Technology of The University of Tokyo (Tokyo, JAPAN). Experiments have been executed in a corridor-like setting composed as in Fig.~\ref{fig:experiments-setting}. The central area of $10\times3$ m$^2$ was recorded for the tracking of participants and it was surrounded by two start areas of $12\times 3$ m$^2$ and two buffer zones of 2 m length that allowed participants to reach a stable speed at the measurement area. Participants were asked to wear coloured caps so that trajectories could be automatically recorded with the software \emph{PeTrack}\cite{boltes2013collecting}.

Each experiment was composed of four procedures, to which 54 male students participated. To achieve a more consistent dataset, every procedure was iterated four times. The aim of the whole investigation has been to test the following hypotheses: (Hp1) the increase of flow ratio negatively impacts the speed of pedestrians; (Hp2) the cohesion of dyad members affects their speed; (Hp3) the cohesion of dyads leads to a lower pedestrian flow at a macroscopic level. 

\begin{figure}[t]
\begin{center}
\includegraphics[width=.99\textwidth]{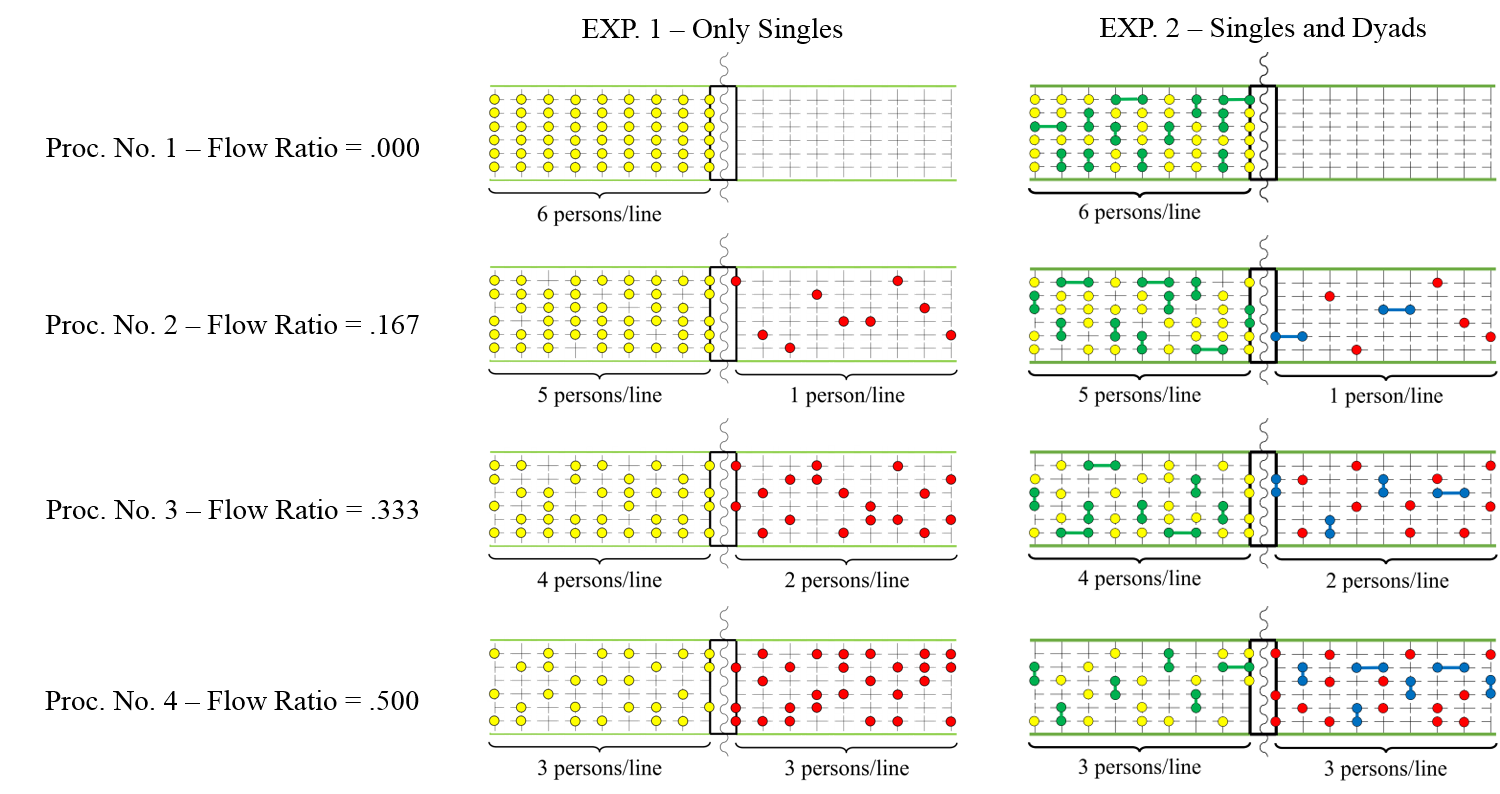}
\caption{Experiments and procedure tested in this investigation.}\label{fig:experiments-procedures}
\end{center}
\end{figure}

With \emph{flow ratio} we denote the rate between the \emph{minor flow} and the \emph{total flow} in bidirectional scenarios. Flow ratio was managed as independent variable among four experimental procedures, as graphically exemplified in Fig.~\ref{fig:experiments-procedures}.

At the beginning of each iteration and according to the tested flow ratio, pedestrians were placed in the marked positions of the start areas. Fig.~\ref{fig:experiments-procedures} exemplifies the arrangement in all experimental procedures. In case of \expr{2}, roughly 44\% of the participants (24 out of the 54 total) was configured as dyads. These were formed by coupling two random members and asking them to possibly walk close to the other companion during the iteration. As shown in Fig.~\ref{fig:experiments-procedures}, dyads could be initially disposed either in \emph{line abreast} or \emph{river-like} pattern, except in \proc{2} where only the latter was possible for dyads belonging to the minor flow. In the other cases the choice was purely random. 

\section{Data Analysis}
Individual data analysis on each of the two experiments has been already described in~\cite{feliciani2016empirical} and~\cite{gorrini2016social}. As expected, both facing a counter-flow and being part of a group was found to influence the walking speed of pedestrians. In addition, there is a difference between the behavior found in balanced and un-balanced configurations of counter-flow in terms of lanes' formation and amount of lateral motion required to avoid conflicts with participants from the opposite direction~\cite{feliciani2016empirical,Feliciani2017new}. In this paper we will compare the results of the two experiments, with the aim of verifying whether their spatial patterns and different speeds affect the dynamics at a more macroscopic level.

\subsection{Microscopic Analysis on Dyads}
\begin{figure}[t]
\begin{center}
\subfigure[]{\includegraphics[width=.48\textwidth]{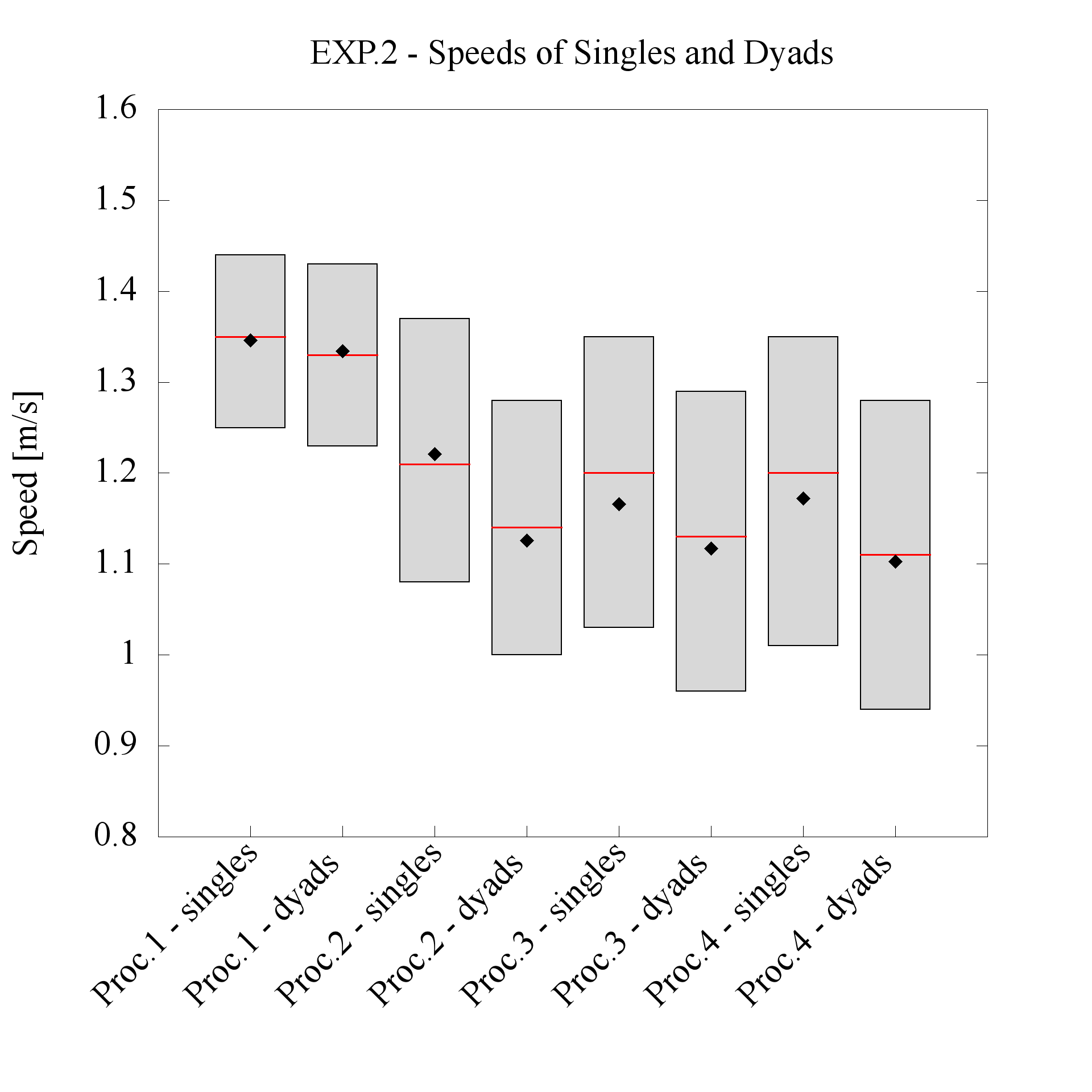}\label{fig:speeds}}
\subfigure[]{\includegraphics[width=.48\textwidth]{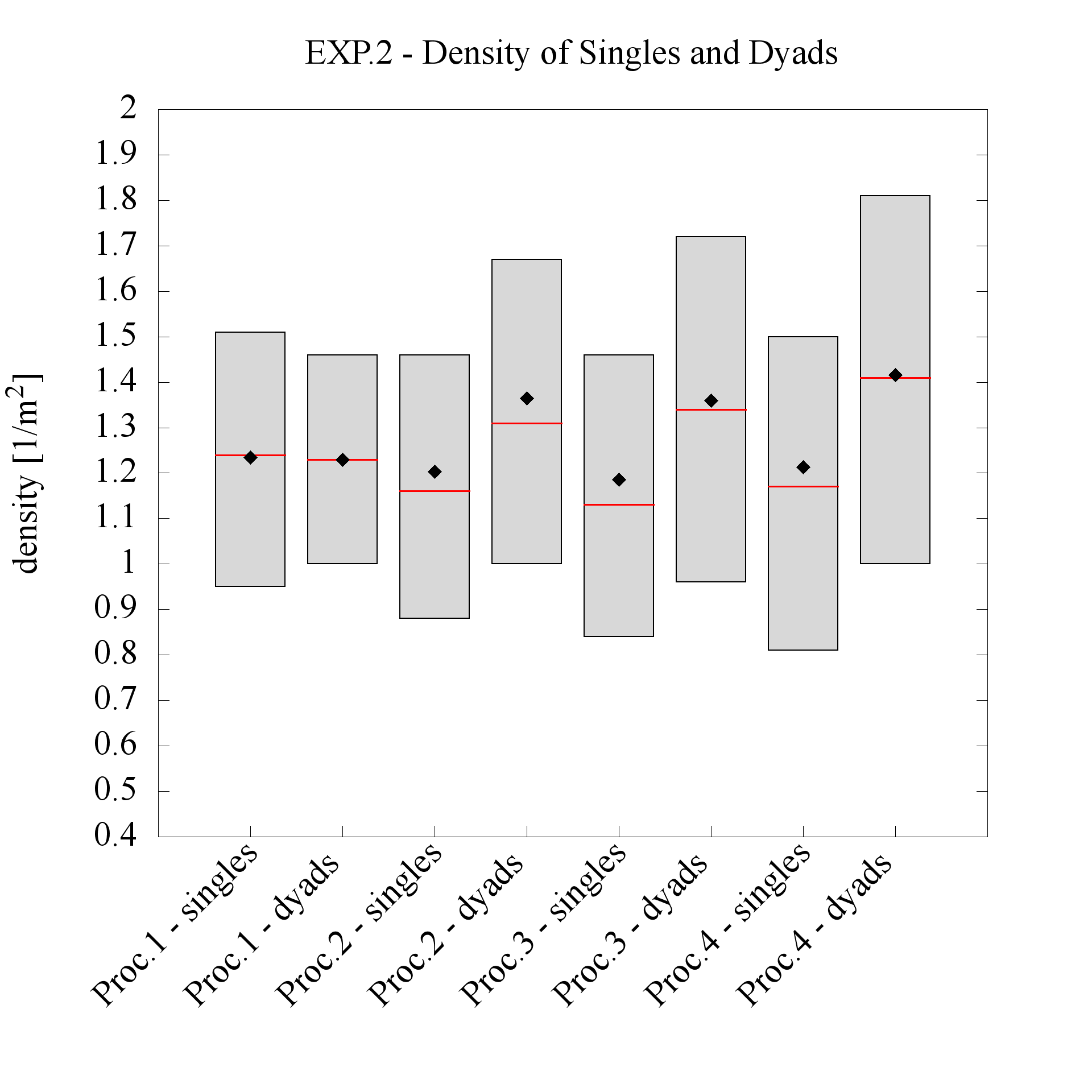}\label{fig:densities}}
\caption{Distribution of pedestrian speeds (a) and local densities (b) among procedures of \expr{2}. Black dots indicate the mean, red lines the median and the box size defines the standard deviation.}
\end{center}
\end{figure}

A comparison of speeds among dyads and individuals in \expr{2} shows that dyads are slower in procedures characterised by a counter-flow situation. In presence of a uni-directional flow essentially exempt from collisions (\proc{1}), on the other hand, the difference is rather small (see Fig.~\ref{fig:speeds}). This suggests that the bi-directional flow affects more the spatial pattern of the dyad members, that more frequently switch from the desired \emph{line-abreast} pattern to a \emph{river-like} one. Moreover, it is also observed that group members have perceived a sensibly higher density during the procedures with counter-flow: Fig.~\ref{fig:densities} shows the distributions of local densities for all procedures of \expr{2}, calculated using the \emph{Voronoi} method~\cite{tordeux2015quantitative} (density values used here are instantaneous and collected from the time the first participant enters the measurement area to the time the last one leaves it). While average density is almost equal in \proc{1}, the difference becomes already noticeable in the second one. Later on we will show that this is due to the fact that dyad members tend to walk close to each other compared to the other individual pedestrians, that instead take more frequently the opportunities offered by unoccupied gaps in front of them.

\begin{figure}[t]
\begin{center}
\subfigure[]{\includegraphics[width=.45\textwidth]{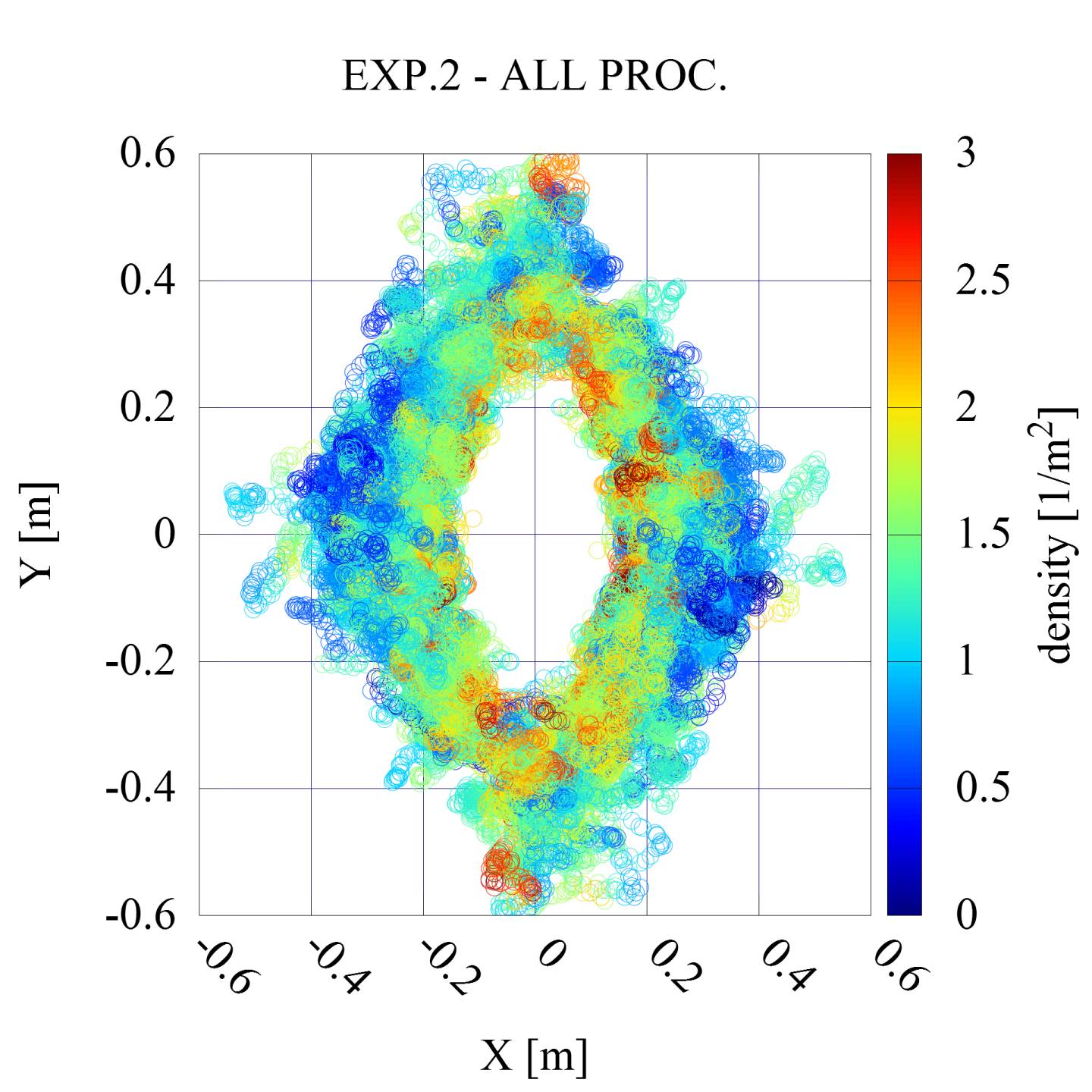}}
\subfigure[]{\includegraphics[width=.45\textwidth]{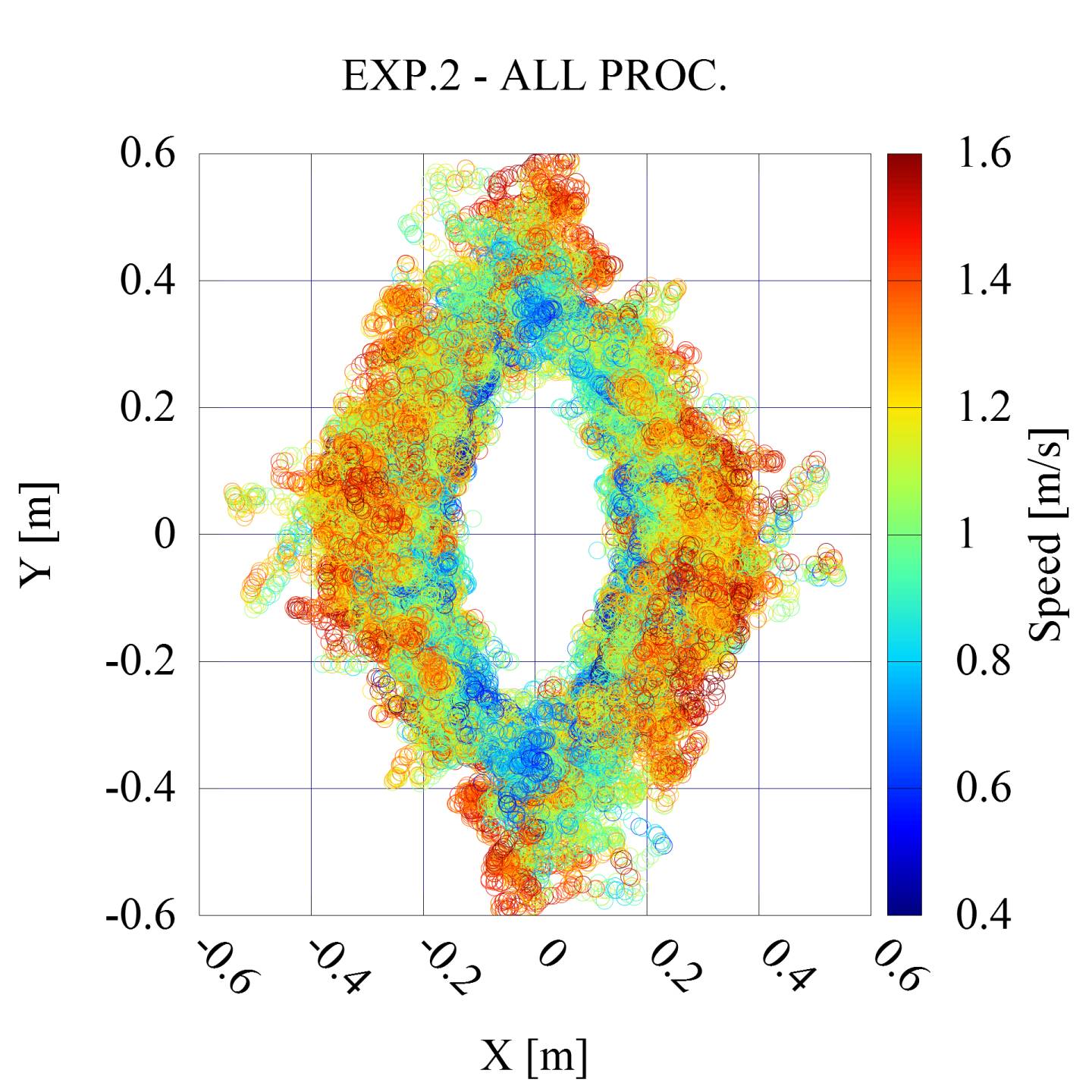}}
\subfigure[]{\includegraphics[width=.45\textwidth]{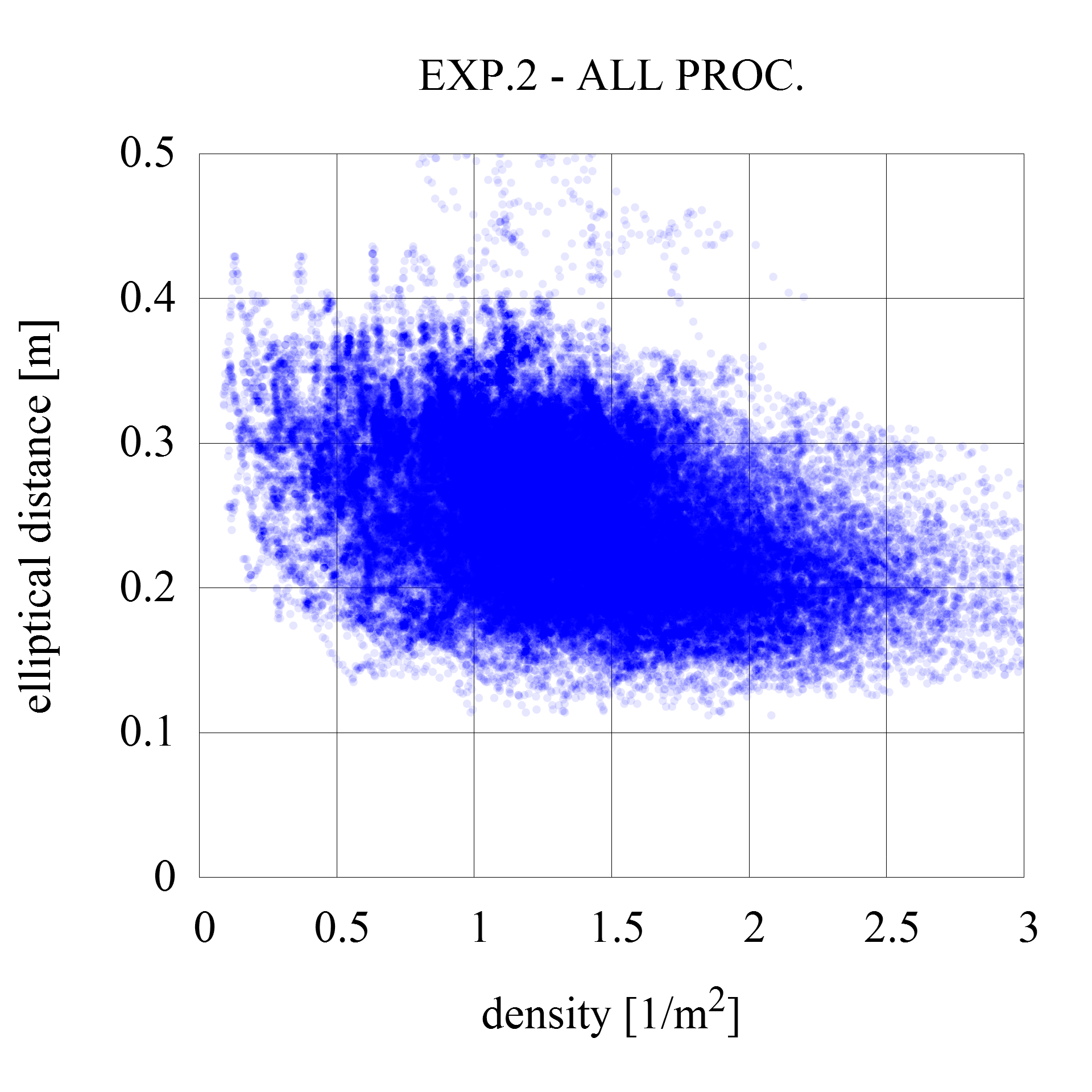}}
\subfigure[]{\includegraphics[width=.45\textwidth]{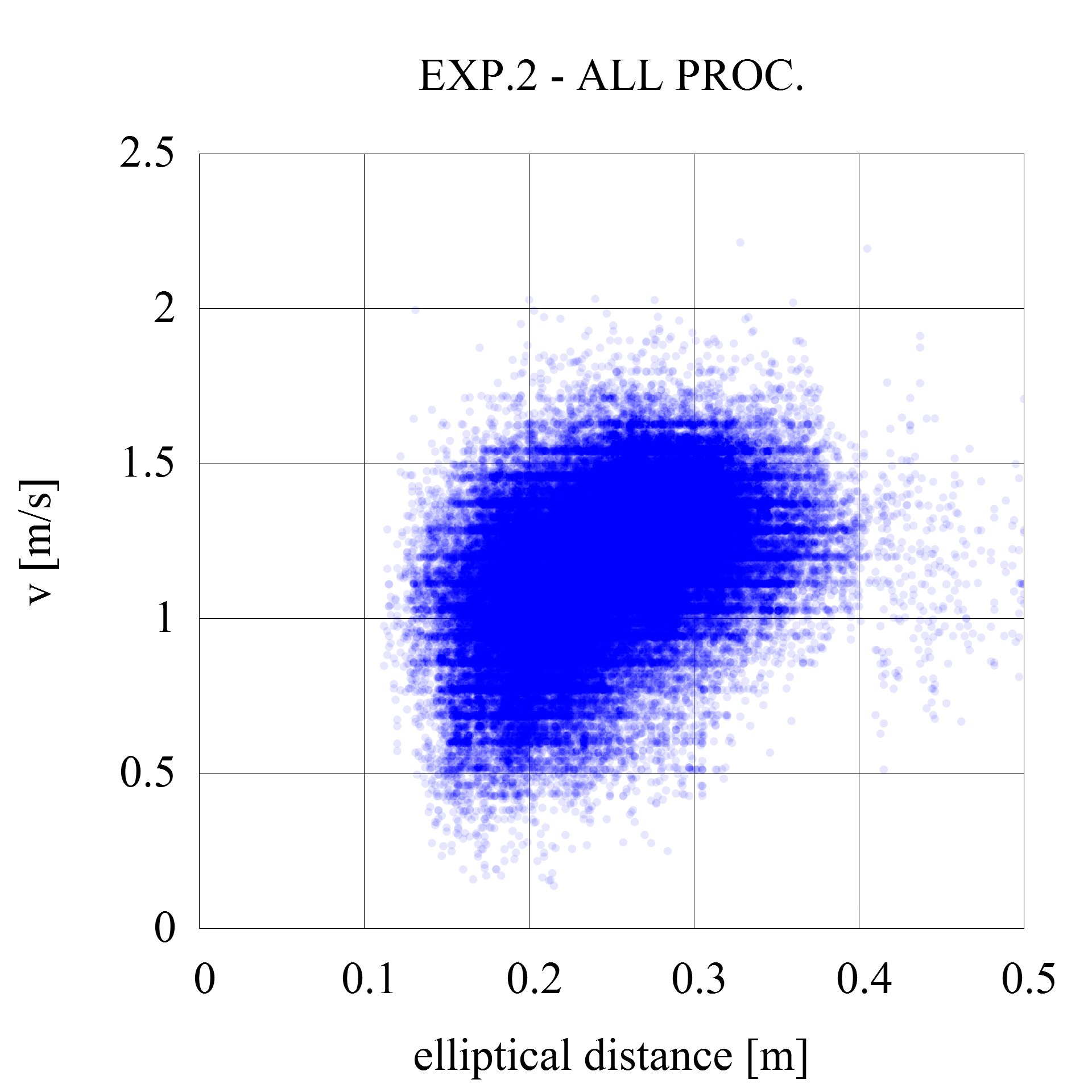}}
\caption{(a -- b) Relative position of dyads according to their centroid. Positions are rotated so that the movement direction is up. Colors indicate the information of local density (a) or instantaneous speed (b). (c -- d) Relations between density, distance and speed of dyads.}\label{fig:dyads_densitySpeed}
\end{center}
\end{figure}

A first analysis on the distributions of relative positions of dyads, with respect to their centroid, has shown a decrease of distance between them with the increase of counter-flow conditions~\cite{gorrini2016social}. Moreover, it is observed that conflicts arising from the bi-directional flow frequently lead dyad members to assume a river-like pattern, which is barely visible in the first procedure. As with the analysis on speed and local density distribution, no significant difference arises between \proc{3 and 4} of the second experiment. The relation between density, speed and relative positions is shown in Fig.~\ref{fig:dyads_densitySpeed}(a) and (b). While a dependency between density and angular arrangement (i.e.~spatial pattern) of dyads was not found to be significant, it is apparent how points of high densities are mostly close to the center (the few outliers are probably due to a transient stretched river-like pattern) and they describe a pattern with an elliptical shape, whose long side is associated to the walking direction. The same regularity is also visible with the data about the speed: points associated to higher speeds are located in the outer part of the dataset, while close to the center speeds are lower, about 0.6 m/s. 

The elliptical appearance of diagrams in Fig.~\ref{fig:dyads_densitySpeed}(a) and (b) is not surprising and it reflects the physics of pedestrian movement already considered with former works on the modelling side (e.g.~\cite{Chraibi2010}). According to these data, it is possible to define a distance metric that applies distortion on the y-axis and helps to analyse the relation between density, speed and \emph{elliptical distance}:

$$f(x,y) = \sqrt{x^2 + \left(\frac{y}{2}\right)^2} $$

The outcome of this analysis is shown in Fig.~\ref{fig:dyads_densitySpeed} (c) and (d). It is fair to state that the defined elliptical distance between dyad members acts as a mediator between the fundamental characteristics of the dynamics (more logically the local density leads group members to walk closer and not vice-versa). On the other hand, this analysis suggests that there is a positive effect of the density on the cohesion of dyad members, that consequently affects their instantaneous speed. However, the relation between walking speed and elliptical distance is less clear and points in Fig.~\ref{fig:dyads_densitySpeed}~(d) appear to be in a rather large area which is difficult to describe using a linear function. Considering these observations, we can say that models of dyads should be able to reproduce a growing trend between elliptical distance and speed. 

\begin{figure}[t]
\begin{center}
\subfigure[]{\includegraphics[width=.45\textwidth]{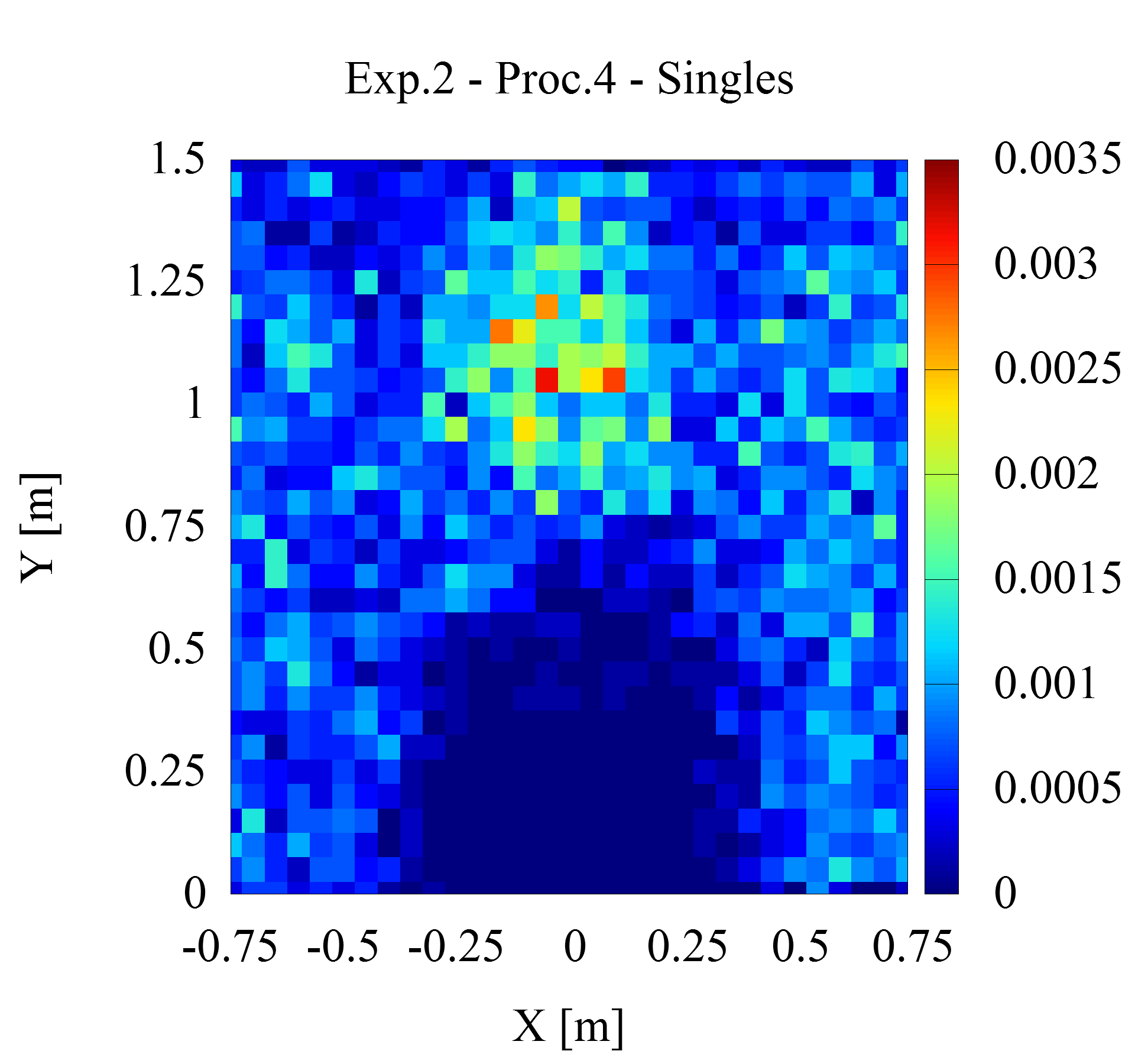}\label{fig:lane_singles}}
\subfigure[]{\includegraphics[width=.45\textwidth]{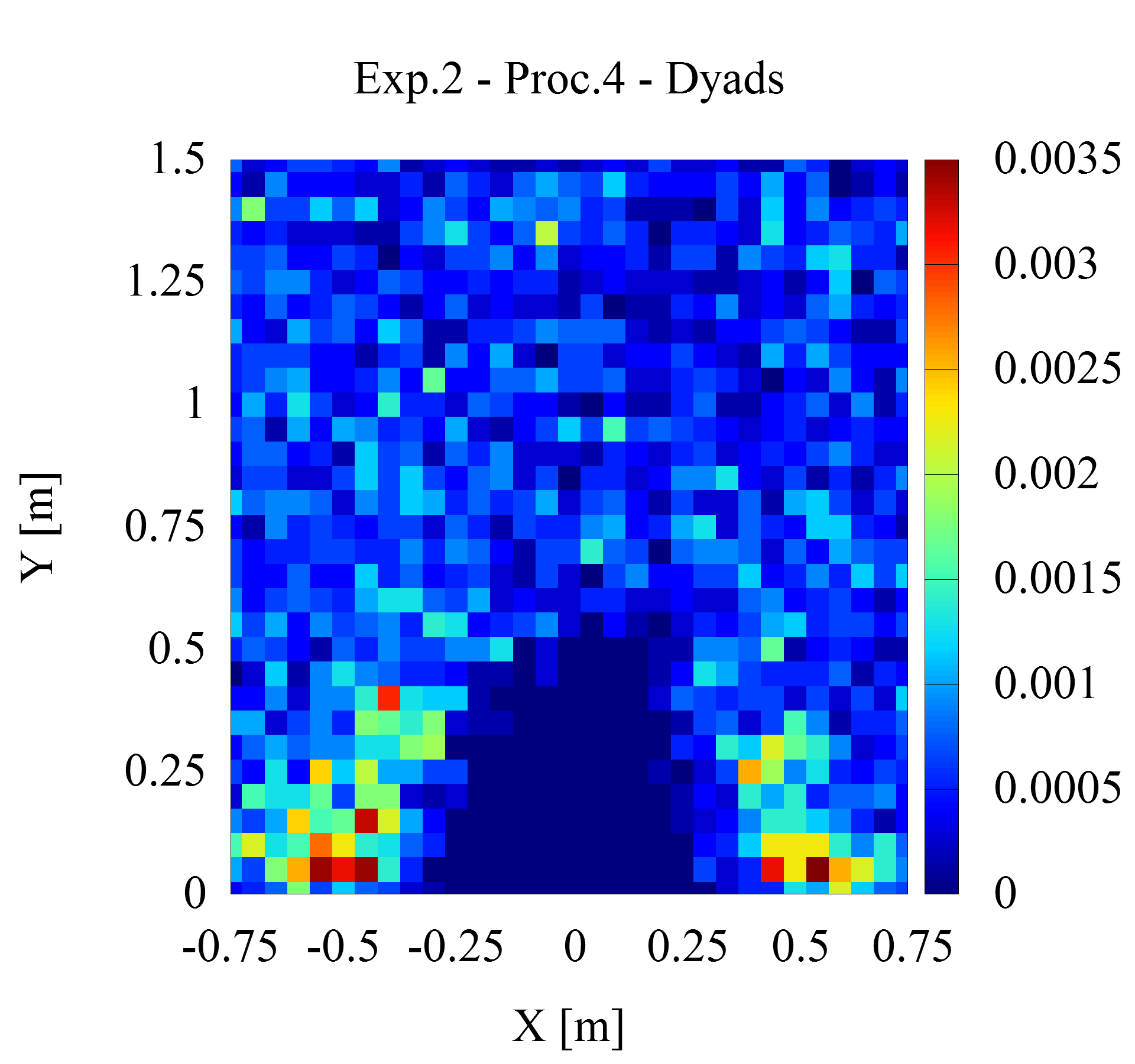}\label{fig:lane_dyads}}
\caption{Distribution of relative positions of neighbour pedestrians, according to the position of each individual (a) or dyad member (b).}\label{fig:lane_formation}
\end{center}
\end{figure}

An additional analysis carried out on microscopic data about the instantaneous position of pedestrians is focused on evaluating the position of the other nearby pedestrians: this kind of analysis, shown in Fig.~\ref{fig:lane_formation}, highlights a different kind of behaviour between members of dyads and individuals with regard to the lane formation phenomenon. We focus in particular on \proc{4} since it is the most interesting one in terms of macroscopic results. It must be said that lane formation is a rather fuzzy concept and several methods are proposed in the literature as attempt for its quantification: \cite{feliciani2016empirical}, for example, analyse the \emph{rotation} of the pedestrian directions to achieve a numerical value describing the stability of lanes. We also do not try to provide a definition of lane, but the data describing the proxemic behaviour of individuals and dyads, respectively shown in Fig.~\ref{fig:lane_formation}(a) and (b), show that, on one side, there is a clear following behaviour for the individuals, where the most frequent position of neighbouring pedestrians is in a spot about 1 m ahead. On the other hand, members of dyads mostly try to keep a line-abreast pattern: the most frequent positions for neighbours are in fact on the side instead that ahead the considered pedestrian. In other words, lanes composed of dyad members tended to be wider and this led to a less efficient utilisation of the space available on average. 


\subsection{Effects of dyads at a macroscopic scale}\label{sec:lane}
Previous results highlighted the effects of density and counter-flow situations on the behaviour of dyads at a very detailed scale whereas we present here the aggregated effect of these microscopic observations on the overall pedestrian flow at different levels of density. 

\begin{figure}[t]
\begin{center}
\subfigure{\includegraphics[width=.45\textwidth]{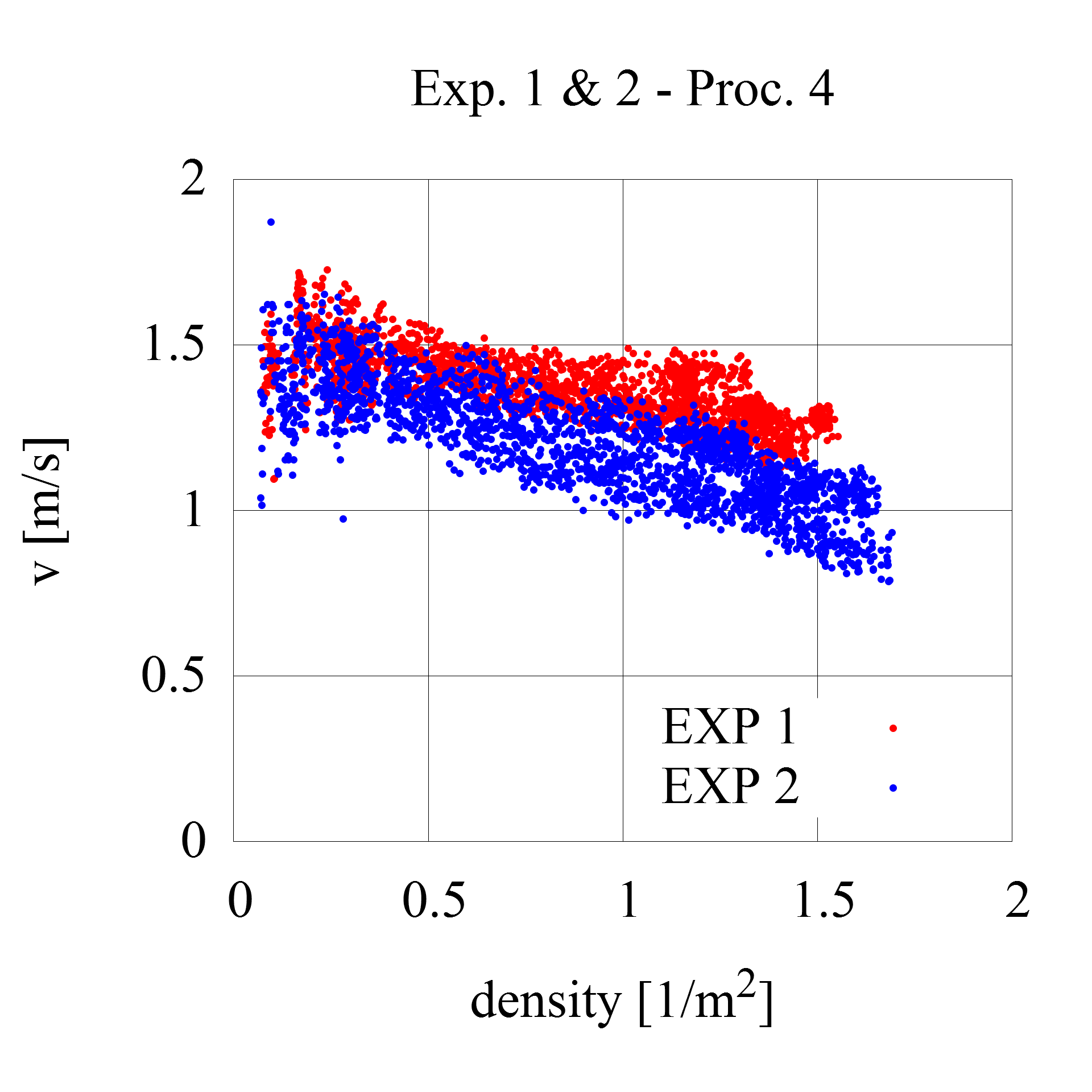}}
\subfigure{\includegraphics[width=.45\textwidth]{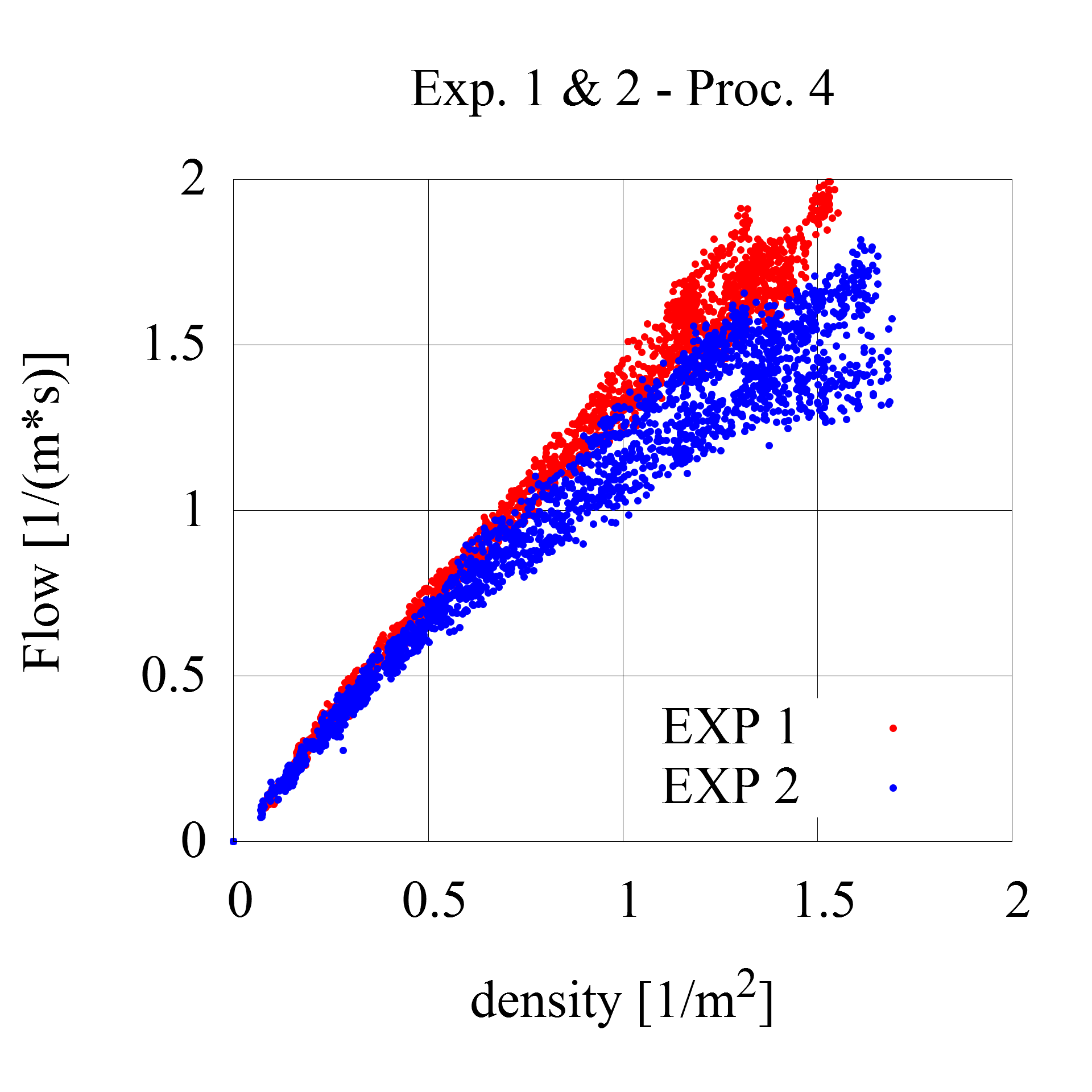}}
\caption{Comparison of fundamental diagrams in the form density--speed (left) and density--flow (right) of \expr{1 and 2} - \proc{4}.}\label{fig:FD}
\end{center}
\end{figure}

The presence of groups in \proc{1} did not bring to significant differences, since only a simple free-flow situation emerged from it. In counter-flow situations, instead, differences become apparent and the most interesting result is represented by the scenario with a perfectly balanced counter-flow, whose data are reported in Fig.~\ref{fig:FD}. The diagram shows very little difference at low densities, but starting from 0.5 peds/m$^2$ the specific pedestrian flow observed in the experiment with dyads grows at a slower rate, compared to the \expr{1}. The range of observed densities does not reach a critical density for any of the experiments, but the trend of the diagrams supports the conjecture that the situation of \expr{2} would lead to a lower maximum flow.

\section{Conclusions and Future Works}

The paper has presented original results of analyses of pedestrian dynamics achieved through an experimental observation aimed at characterising the influence of dyads, both at micro and macroscopic level. Micro-level results underline that different counterflow situations affect local density, and that groups walk slower compared to singletons, depending also on their spatial patterns at variable density situations. The introduction of dyads in the pedestrian demand leads to a higher level of measurable density in analogous initial conditions and a more chaotic macro-level dynamics characterized by fragmented lanes, inducing a lower observed specific flow. 

Future works are aimed, on one hand, to transfer the achieved results to the modelling activities in presence of groups (preliminary results are discussed in another paper in this volume~\cite{TGF2017-Yiping}), but additional observations and experiments would be needed to further investigate whether previously observed aggregated phenomena are still observed in the presence of groups.

\end{document}